\documentclass[aps,prb,twocolumn]{revtex4} 
\usepackage{amssymb}
\usepackage{graphicx} 
\usepackage{amsmath}
\usepackage{color}

\providecommand{\vect}[1]{{\boldsymbol{#1}}}

\usepackage{physics}

\begin{document} 
\title{Spin pumping in noncollinear antiferromagnets} 

\author{Mike A. Lund,  Akshaykumar Salimath, and Kjetil M. D. Hals} 
\affiliation{Department of Engineering Sciences, University of Agder, 4879 Grimstad, Norway} 
\newcommand{\Kjetil}[1]{\textcolor{red}{#1}} 
\begin{abstract}
The ac spin pumping of noncollinear antiferromagnets is theoretically investigated. Starting from an effective action description of the spin system, we derive the Onsager coefficients connecting the spin pumping and spin-transfer torque associated with the dynamics of the SO(3)-valued antiferromagnetic order parameter. Our theory is applied to a kagome antiferromagnet resonantly driven by a uniform external magnetic field. We demonstrate that the reactive (dissipative) spin-transfer torque parameter can be extracted from the pumped ac spin current in-phase (in quadrature) with the driving field. 
Furthermore, we find that the three spin-wave bands of the kagome AF generate spin currents with mutually orthogonal polarization directions. This offers a unique way of controlling the spin orientation of the pumped spin current by exciting different spin-wave modes. 

\end{abstract}

\maketitle 

\section{Introduction} 
Over the last years, there has been rapidly growing interest in implementing antiferromagnetic elements in spin-based electronics. This has led to the development of antiferromagnetic spintronics~\cite{Jungwirth:np2018,Duine:np2018,Gomonay:np2018,Zelezny:np2018,Nemec:np2018,Libor:np2018}, in which the information is coded into the magnetic moments of  antiferromagnets (AFs)~\cite{Neel:AnnPhys1967}. 
Unlike ferromagnets, which have been the traditional building blocks of spintronics, the AFs are remarkably stable against magnetic field noise due to their vanishing magnetization.  
Furthermore, the AFs are characterized by terahertz (THz) spin dynamics, which is a thousand times faster than the characteristic frequency of ferromagnets. The ultrahigh-frequency of AFs is desirable for use in future spin electronics because it allows for significantly higher operational speeds of the devices. Ultrafast switching of AFs has been experimentally demonstrated~\cite{Wadley:science2016}, and recent works have shown that 
the antiferromagnetic order is efficiently manipulatable via electric currents~\cite{Wadley:science2016,Reichlova:prb2015,Nunez:prb2006,Duine:prb2007,Gomonay:jmj2008,Wang:prl2008,Haney:prl2008,Gomonay:prb2010,Hals:prl2011,Gomonay:prb2012,Manchon:prb2014,Cheng:prb2014,Cheng:prl2014,Velkov:njp2016} as well as optical pulses~\cite{Duong:prl2004,Kimel:n2004,Manz:np2016}.

A cornerstone of spintronics is the ability to manipulate the order parameter of magnetic materials via spin-transfer torques (STTs) -- the process where spin currents produce magnetic torques via direct transfer of spin angular momentum from the itinerant electrons to the ordered spin system~\cite{Ralph:JMMM2008}. The reciprocal process of the STT is spin pumping and refers to the phenomenon where the collective spin excitations of the magnet pump a spin current into adjacent metallic leads~\cite{Tserkovnyak:rmp2005}. Notably, the linear response coefficients describing the STT and spin pumping are connected via the Onsager reciprocal relations~\cite{Groot:1952, Hals:EPL2010,Hals:prb13,Hals:prb15}.  Consequently, one can obtain significant insight into the strength, symmetry, and governing mechanisms of the STT by probing the reciprocal spin pumping process.  

\begin{figure}[ht] 
\centering 
\includegraphics[scale=1.0]{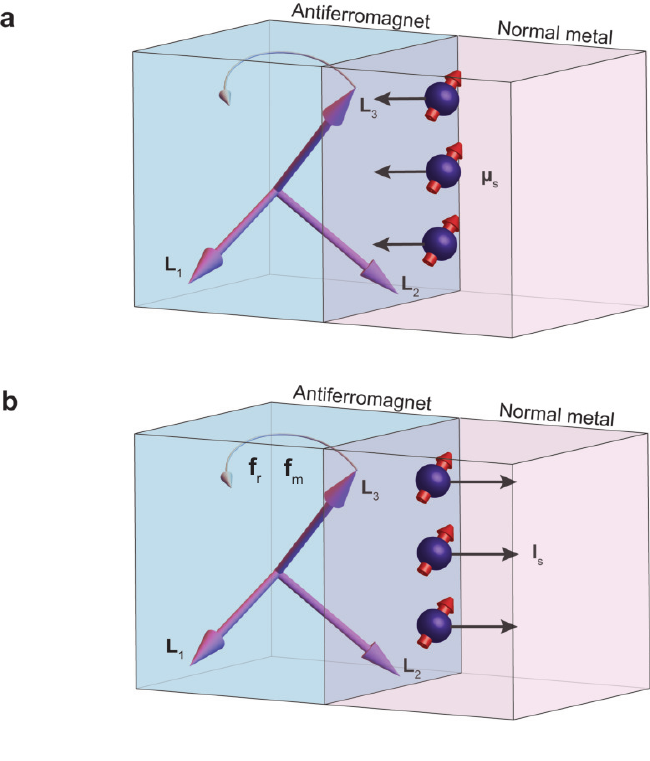}  
\caption{(color online). 
Illustration of two reciprocal spin processes in a bilayer consisting of an NCAF interfaced with a normal metal (NM).
 {\bf a}. A spin accumulation $\boldsymbol{\mu}_{s}$ in the NM induces a spin current $\boldsymbol{I}_s$ that flows into the NCAF, producing an STT on the NCAF.  
 {\bf b}. A uniformly precessing NCAF, which is driven by the effective fields $\boldsymbol{f}_r$ and $\boldsymbol{f}_m$, pumps a spin current $\boldsymbol{I}_{s}$ into the NM layer. For a uniformly precessing NCAF with strong exchange interaction, the antiferromagnetic spin state can be parametrized by the macroscopic staggered vectors $\boldsymbol{L}_1$, $\boldsymbol{L}_2$, $\boldsymbol{L}_3$, and the dynamics can be described as a rotation of the coordinate system spanned by these vectors. The STT and spin pumping in {\bf a} and {\bf b}, respectively, are connected via the Onsager reciprocal relations.  
 }
\label{Fig1} 
\end{figure} 

In AFs, the spin pumping~\cite{Cheng:prl2014,Kamra:prl2017,Troncoso:prb2021} and STT~\cite{Nunez:prb2006,Gomonay:prb2010,Gomonay:prb2012} have been theoretically investigated in several works, and two experiments recently observed sub-THz spin pumping in the uniaxial insulating AFs MnF$_2$~\cite{Vaidya:Science2020} and Cr$_2$O$_3$~\cite{Li:Nature2020}. Most of these works have concentrated on collinear AFs, which are antiferromagnetic systems characterized by a single order parameter (commonly known as the staggered field or N\'eel vector). However, several AFs require two or three mutually orthogonal staggered fields to describe the spin order correctly (see Fig.~\ref{Fig1})~\cite{Andreev:spu1980}. In this case, the system is referred to as a noncollinear AF (NCAF). The spin order of NCAFs is parameterized by a rotation matrix, which defines the orientation of the reference frame spanned by the orthogonal staggered fields~\cite{Andreev:spu1980,Dombre:prb1989}. 
To date, little knowledge exists on how spin currents couple to the SO(3) order parameter of NCAFs. In Ref.~\onlinecite{Gomonay:prb2012}, the STT was phenomenologically investigated, whereas the dissipative coupling mechanism was derived in Ref.~\onlinecite{Tserkovnyak:prb2017} from a scattering matrix formalism and applied to amorphous magnets and kagome AFs in Refs.~\onlinecite{Ochoa:prb2018,Li:prb2021}. However, so far,  there have been no thorough investigations of the spin pumping process in these nontrivial spin systems. 

In this work, we derive a general theory of the reactive and dissipative ac spin pumping in NCAFs. The general formalism is applied to NCAFs with kagome lattice structure. Importantly, we find that both the reactive and dissipative STT parameters can be mapped out from the spin pumping signal measured via the inverse spin Hall effect (ISHE). Additionally, we show that the three spin-wave bands of the kagome AF produce spin currents with orthogonal spin polarizations, which enables manipulation of the spin current's orientation by only tuning the frequency of the external driving field. When the driving field hits the resonance frequency of a spin-wave band, a current with a fixed spin polarization is created. This phenomenon differs markedly from spin pumping of ferromagnets and collinear AFs, where a reorientation of the magnetic state is required for changing the polarization direction. 
Thus, our work demonstrates that spin pumping could represent an effective technique for exploring novel spin torque mechanisms in NCAFs.

This paper is organized as follows. Sec.~\ref{Sec:Theory} presents a general effective action description of NCAFs and derives the Onsager coefficients representing the coupling between the NCAF and spin currents. From the Onsager coefficients, a general theory of spin pumping is derived. Then, in Sec.~\ref{Sec:Kagome}, the general theory is applied to kagome AFs. A summary is provided in Sec.~\ref{Sec:Conclusions}, whereas the action and dissipation functionals of kagome AFs are microscopically derived in appendixes~\ref{Sec:Appendix} and \ref{Sec:Dissipation}.

\section{General theory}\label{Sec:Theory} 
We consider the reciprocal processes \emph{spin pumping} and \emph{STT} in a bilayer consisting of an NCAF of volume $V$ interfaced with a normal metal (NM) (see Fig.~\ref{Fig1}).~\cite{Hals:EPL2010, Cheng:prl2014, Maekawa:book} 
Our main aim is to derive a general theory for the spin pumping of a \emph{uniformly} precessing NCAF.
To this end, we first consider the STT, derive the Onsager coefficients governing the STT-driven uniformly precessional motion of the NCAF, and then use the Onsager reciprocal relations to find an expression for the spin pumping. 

Our model is based on the assumption that the exchange interaction of the NCAF is much stronger than any other interaction energies in the microscopic spin Hamiltonian such that the mutual orientation of the sublattice spins only is weakly affected by the dynamics of the NCAF.~\cite{Andreev:spu1980,Dombre:prb1989} The STT is produced by a spin accumulation $\boldsymbol{\mu_{s}}$ in the NM layer at the NM/NCAF interface.~\cite{Maekawa:book} 
The vector $\boldsymbol{\mu_{s}}$ has a direction parallel to the out-of-equilibrium spin density in the NM and a norm equal to the difference between the chemical potentials of the spin up and down electrons.
The spin accumulation yields a spin current that flows into the NCAF, transferring its spin angular momentum to the antiferromagnetic system. The source of this spin accumulation does not play a role in the theory we develop and could, in principle, originates from any microscopic mechanism generating an out-of-equilibrium spin density (e.g., spin Hall effect). Furthermore, we disregard the effects of the SOC that break the spin rotational symmetry of the STT.  
In this case, the effective action $\mathcal{S}$ of the NCAF can to second order in the space-time gradients and external forces (i.e., STTs and magnetic fields) be written as~\cite{Dombre:prb1989}
\begin{equation}
    \mathcal{S} = \int {\rm d}V{\rm d}t \mathcal{L}.\label{Eq:Sgeneral}
\end{equation}
The Lagrangian density $\mathcal{L}= \mathcal{T} - \mathcal{U} - \mathcal{U}_s$ of the spin system consists of a kinetic term $\mathcal{T}$
\begin{equation}
    \mathcal{T} = \frac{a_1}{2} \boldsymbol{\mathcal{V}}\cdot\boldsymbol{m},\label{Eq:Tgeneral}
\end{equation}
the energy $\mathcal{U}$ produced by the exchange interaction, the spin-orbit coupling (SOC), and the magnetic field $\sim\boldsymbol{h}$ 
\begin{equation}
    \mathcal{U} = \Lambda_{ij}^{\alpha\beta} \left[ \partial_{\alpha} \mathbf{R}^T  \partial_{\beta} \mathbf{R} \right]_{ij} + \nu_{ij}^{kl} R_{ij} R_{kl}  + \tilde{\kappa}_{ij}m_i m_j   - \boldsymbol{h}\cdot\boldsymbol{m}  ,  \label{Eq:Ugeneral}
\end{equation}
and a term $\mathcal{U}_s$ representing the coupling to the spin accumulation $\boldsymbol{\mu}_s$ of the itinerant quasi-particles that diffuse into the NCAF from the adjacent NM layer
\begin{equation}
    \mathcal{U}_s = \lambda \boldsymbol{m}\cdot \boldsymbol{f}_s . \label{Eq:Usgeneral}
\end{equation}
Here, $a_1$ is a constant that depends on the lattice structure, $\boldsymbol{R}$ is a rotation matrix that describes the orientation of the reference frame spanned by the staggered fields of the NCAF, $\mathcal{V}_i= -(1/2)\epsilon_{ijk} [\boldsymbol{R}^T \dot{\boldsymbol{R}}]_{jk}$ (where $\epsilon_{ijk}$ is the Leivi-Civita tensor and $\dot{\boldsymbol{R}} \equiv \partial_t \boldsymbol{R} $), $\boldsymbol{f}_s= \boldsymbol{\mu}_s/\hbar$, and 
$\boldsymbol{m}$ is proportional to the out-of-equilibrium magnetization produced by a relative tilting of the magnetic sub-lattices. 
$\partial_{\alpha} \mathbf{R}$ represents the partial derivatives of the rotation matrix with respect to the spatial coordinates, i.e., $\alpha  \in \{ x,y,z\}$.
For further analysis, it is convenient to split $\tilde{\kappa}_{ij}$ into isotropic and anisotropic terms: $\tilde{\kappa}_{ij}= a_2\delta_{ij} + \eta_{ij}$ ($\delta_{ij}$ is the Kronecker delta).
The coefficients $\Lambda_ {ij}^{\alpha\beta}$ and $a_2$ ($\nu_{ij}^{kl}$ and $\eta_{ij}$) are proportional to the isotropic exchange interaction (the anisotropy energy), 
whereas the constant $\lambda$ parameterizes the strength of the reactive STT induced by the spin accumulation.
Throughout, Einstein’s summation convention is implied for repeated indices. 

The dissipative processes in the NCAF is determined by the dissipation functional~\cite{Gomonay:prb2012,Rodrigues:arxiv21}
\begin{eqnarray}
\mathcal{G} &=& \int {\rm d}V{\rm d}t \left[ \frac{\tilde{\alpha}}{8} {\rm Tr}\left(  \dot{ \mathbf{R}}^T \dot{ \mathbf{R} }\right) + \frac{\tilde{\lambda}}{2} \boldsymbol{\mathcal{V}}\cdot \boldsymbol{f}_s  \right] , \label{Eq:Dissipation} 
\end{eqnarray}
where $\tilde{\alpha}$ and $\tilde{\lambda}$ parameterize the damping of the spin system and the dissipative STT, respectively.  

In Sec.~\ref{Sec:Appendix}-\ref{Sec:Dissipation}, we microscopically derive the above action and dissipation functional for a NCAF with a kagome lattice. 
For the sake of completeness, we have in Eq.~\eqref{Eq:Ugeneral} also included the energy contribution from the spatial variations of the order parameter.
However, when deriving the Onsager coefficients, we will restrict ourselves to the uniform case and disregard the gradient terms. 

We consider small deviations from the uniform equilibrium state of the NCAF.
In this case, it is convenient to use a Gibbs vector representation of the SO(3) rotation matrix $\boldsymbol{R}$.
The Gibbs vector $\boldsymbol{r}$ corresponding to a rotation by an angle of $\theta$ about the axis $\hat{\boldsymbol{n}}$ is $\boldsymbol{r}= \tan (\theta/2) \hat{\boldsymbol{n}}$,   
and the action of $\boldsymbol{R}$ on a general vector $\boldsymbol{v}$ is~\cite{Haslwanter:book, Andreev:spu1980}  
\begin{eqnarray}
\boldsymbol{R}\boldsymbol{v} &=& \boldsymbol{v} + \frac{2}{1+|\boldsymbol{r}|^2}\left[ \boldsymbol{r}\times\boldsymbol{v} + \boldsymbol{r}\times (  \boldsymbol{r} \times \boldsymbol{v} )  \right] . \label{Eq:GibbsRel1}
\end{eqnarray}
Furthermore, it is possible to represent the partial derivatives $\partial_{\mu} \boldsymbol{R}$ of the rotation matrix in terms of the Gibbs vector via the relationship (here, $\mu\in\{t,x,y,z\}$)~\cite{Haslwanter:book} 
 \begin{eqnarray}
\left[(\partial_{\mu} \boldsymbol{R})\boldsymbol{R}^{T}\right]_{ij} &=& \epsilon_{ikj} \frac{2}{1+|\boldsymbol{r}|^2}\left[ \partial_{\mu} \boldsymbol{r} + \boldsymbol{r}\times\partial_{\mu} \boldsymbol{r}  \right]_k . \label{Eq:GibbsRel2}
\end{eqnarray}   
Note that the identity matrix corresponds to $\boldsymbol{r}= \boldsymbol{0}$.
Thus, $|\boldsymbol{r}| \ll 1 $ since we consider small deviations from the equilibrium state. 
Using Eqs.~\eqref{Eq:GibbsRel1}-\eqref{Eq:GibbsRel2} and keeping terms in Eqs.~\eqref{Eq:Tgeneral}-\eqref{Eq:Usgeneral} up to second order in the out-of-equilibrium quantities $\{ \boldsymbol{r}, \boldsymbol{m} \}$ and external force fields $\{ \boldsymbol{f}_s, \boldsymbol{h} \}$,  we find the Lagrange density 
\begin{eqnarray}
\mathcal{L} &=& a_{1} \boldsymbol{m}\cdot\dot{\boldsymbol{r}} - \mathcal{U} (\boldsymbol{r},\boldsymbol{m}) - \mathcal{U}_s (\boldsymbol{m},\boldsymbol{f}_s)   . \label{Eq:Lgeneral}
\end{eqnarray} 
Here,  $\mathcal{U}_s  (\boldsymbol{m},\boldsymbol{f}_s) $ is given by Eq.~\eqref{Eq:Usgeneral}, and the potential $\mathcal{U} (\boldsymbol{r},\boldsymbol{m})$ takes the form of 
\begin{equation}
\mathcal{U} = \Gamma_{ij}^{\alpha\beta} \partial_{\alpha} r_i\partial_{\beta} r_j  +  \kappa_{ij} r_i r_j + \tilde{\kappa}_{ij} m_i m_j  - \boldsymbol{h}\cdot\boldsymbol{m} ,
\end{equation} 
where we have introduced the anisotropy tensor 
$\kappa_{ij}= 2\epsilon_{kil} [2\epsilon_{mjn}\nu_{kl}^{mn} + \epsilon_{ljm} (\nu_{nn}^{km} + \nu_{km}^{nn}) ]$ 
and the exchange tensor $\Gamma_{ij}^{\alpha\beta}= 4[  \Lambda_{kk}^{\alpha\beta}\delta_{ij} - \Lambda_{ij}^{\alpha\beta}]$.
The dissipation~\eqref{Eq:Dissipation} becomes 
\begin{eqnarray}
\mathcal{G} &=& \int {\rm d}V{\rm d}t \left[ \tilde{\alpha} \dot{\boldsymbol{r}}^2  + \tilde{\lambda} \dot{\boldsymbol{r}}\cdot \boldsymbol{f}_s  \right] . \label{Eq:Dissipation2} 
\end{eqnarray}

Eqs.~\eqref{Eq:Lgeneral} and \eqref{Eq:Dissipation2} provide a general effective description of the NCAFs' dynamics.
In absence of external force fields, the Lagrangian \eqref{Eq:Lgeneral} is equivalent to the phenomenological theory derived from symmetry arguments in Ref.~\onlinecite{Andreev:spu1980}. 
This can be seen by minimizing the action with respect to $\boldsymbol{m}$, which yields $\boldsymbol{m} =  (a_1/2a_2) [\boldsymbol{I} + \boldsymbol{\eta}/a_2]^{-1}\dot{\boldsymbol{r}}$ ($\boldsymbol{I}$ is the identity matrix).
The norm of the matrix $\boldsymbol{\eta}/a_2$ is small, because $\boldsymbol{\eta}$ is proportional to the SOC whereas $a_2$ is linear in the strong antiferromagnetic exchange interaction.
This implies that $[\boldsymbol{I} + \boldsymbol{\eta}/a_2]^{-1} \approx [\boldsymbol{I} - \boldsymbol{\eta}/a_2]$. Substituting the expression for $\boldsymbol{m}$ back into the
Lagrange density leads to
\begin{equation}
\mathcal{L} = \chi_{ij}\dot{r}_i\dot{r}_j - \Gamma_{ij}^{\alpha\beta} \partial_{\alpha} r_i\partial_{\beta} r_j - \kappa_{ij} r_i r_j,\label{Eq:NonLinSigma}
\end{equation}  
where $\chi_{ij}= (a_1^2/4a_2)[\delta_{ij} - \eta_{ij}/a_2]$. To second order in $\boldsymbol{r}$, Eq.~\eqref{Eq:NonLinSigma} is identical to the phenomenology developed in Ref.~\onlinecite{Andreev:spu1980}.
The STT-induced coupling terms in Eqs.~\eqref{Eq:Usgeneral} and \eqref{Eq:Dissipation} was phenomenologically derived in Ref.~\onlinecite{Gomonay:prb2012} based on the spin conservation principle. 
 
Next, we will use Eqs.~\eqref{Eq:Lgeneral}-\eqref{Eq:Dissipation2} to derive a general expression for the spin pumping in NCAFs.
We consider a spatial uniform driving field $\boldsymbol{h} (t)$ such that the spatial variations of $\boldsymbol{r}$ and $\boldsymbol{m}$ can be disregarded.
The NM is assumed to act as a perfect spin sink, implying that the backflow of spin from the NM to the NCAF is negligible.

Generally, the state of a system can be described by a set of thermodynamic variables $\{ q_i | i=1,2,3 ... \}$. Let $f_i$ denote the thermodynamic force that induces a flux $J_i$ in the quantity $q_i$. 
In linear response, the fluxes are given by the equation $J_i= L_{ij}f_j$, where the off-diagonal elements of the response matrix $[ L_{ij} ]$ are related via the Onsager reciprocal relations 
$L_{ij}= \epsilon_i\epsilon_j L_{ji}$.  Here, $\epsilon_i= 1$ ($\epsilon_i= -1$) if the thermodynamic variable $q_i$ is even (odd) under time reversal~\cite{Groot:1952}. 
At constant temperature $T$, the fluxes and forces are chosen such that the entropy generation $\mathbb{S}$ is given by $T\dot{\mathbb{S}}=\sum_i J_i f_i$.~\cite{Groot:1952}
NCAFs are described by the variables $\boldsymbol{r}$ and $\boldsymbol{m}$, which under time reversal transform as $\boldsymbol{r}\mapsto\boldsymbol{r}$ and $\boldsymbol{m}\mapsto -\boldsymbol{m}$. 
The fluxes of the NCAF are $\dot{\boldsymbol{r}}$ and $\dot{\boldsymbol{m}}$, whereas the associated forces are $\boldsymbol{f}_r= - V\partial_{\boldsymbol{r}}\mathcal{U}$ and $\boldsymbol{f}_m= - V\partial_{\boldsymbol{m}}\mathcal{U}$, respectively.~\cite{Hals:EPL2010,Hals:prb13,Hals:prb15}
Note that we consider a uniformly precessing NCAF. Thus, $\boldsymbol{r}$ and $\boldsymbol{m}$ parametrize the uniform spin state of the entire NCAF.
During precessional motion of the isolated NCAF, the heat generation of the antiferromagnetic system is $\boldsymbol{f}_r\cdot\dot{\boldsymbol{r}} + \boldsymbol{f}_m\cdot\dot{\boldsymbol{m}} \sim T\dot{\mathbb{S}}$.
In the NM, $\boldsymbol{f}_s$ is proportional to the out-of-equilibrium spin accumulation, which leads to a flow of spin from the NM into the NCAF.
The heat generation of this process is $\boldsymbol{f}_s\cdot \boldsymbol{I}_s \sim T\dot{\mathbb{S}}$, which implies that $\boldsymbol{f}_s$ is the thermodynamic force producing the pure spin current $\boldsymbol{I}_s$.~\cite{Hals:EPL2010,Maekawa:book} 
Thus, $\boldsymbol{I}_s$ and $\boldsymbol{f}_s$ are the flux and force of the out-of-equilibrium spin density $\boldsymbol{\rho}_s$ at the NM/NCAF interface, 
which under time reversal transforms as $\boldsymbol{\rho}_s \mapsto -\boldsymbol{\rho}_s$.
The relationship between the fluxes and thermodynamic forces are given by the equation    
\begin{equation}
\begin{pmatrix}
\dot{\boldsymbol{r}} \\
\dot{\boldsymbol{m}} \\
\boldsymbol{I}_{s}
\end{pmatrix}
=
\begin{pmatrix}
\boldsymbol{L}_{r r} & \boldsymbol{L}_{r m} & \boldsymbol{L}_{r s} \\
\boldsymbol{L}_{m r} & \boldsymbol{L}_{m m} &  \boldsymbol{L}_{m s} \\
\boldsymbol{L}_{s r} & \boldsymbol{L}_{s m } &  \boldsymbol{L}_{s s }
\end{pmatrix}
\begin{pmatrix}
\boldsymbol{f}_{r}  \\
\boldsymbol{f}_{m} \\
\boldsymbol{f}_{s}
\end{pmatrix} . \label{Eq:OnsagerMatrix}
\end{equation} 
Based on Eqs.~\eqref{Eq:Lgeneral}-\eqref{Eq:Dissipation2}, the Onsager coefficients governing the dynamics of $\boldsymbol{r}$ and $\boldsymbol{m}$ can be derived from the NCAF's equations of motion. From the variational equation $\delta\mathcal{S}/\delta\boldsymbol{q} = \delta\mathcal{G}/\delta\dot{\boldsymbol{q}} $ ($\boldsymbol{q}\in \{\boldsymbol{r}, \boldsymbol{m} \}$), we find 
$L_{r_ir_j}=0$, $L_{r_im_j} =-\delta_{ij}/a_1V$, $L_{r_is_j} =\lambda \delta_{ij}/a_1$,
$L_{m_ir_j}= \delta_{ij}/a_1V$, $ L_{m_im_j}=2\tilde{\alpha}\delta_{ij}/ a_1^2V$, and $L_{m_is_j}=-\tau\delta_{ij}/ a_1 $.
Here, we have defined $\tau= (\tilde{\lambda} a_1 + 2\tilde{\alpha} \lambda)/a_1$.
We see that the off-diagonal elements describing the dynamics of the isolated spin system satisfy the expected reciprocity relations  $L_{r_im_j} = - L_{m_jr_i}$. 
The coefficients $L_{r_is_j}$ and $L_{m_is_j}$ define the STT produced by the spin accumulation. The Onsager reciprocal relations implies that $L_{s_ir_j} = -L_{r_j s_i}$ and  $L_{s_im_j}= L_{m_j s_i}$, which yields the spin pumping 
\begin{equation}
\boldsymbol{I}_s = -\frac{\lambda}{a_1}\boldsymbol{f}_r -\frac{\tau}{a_1}\boldsymbol{f}_m  .\label{Eq:IsGeneral}
\end{equation}
Eq.~\eqref{Eq:IsGeneral} is the first central results of this paper and represents a general theory for the ac spin pumping of NCAFs. 
In Eq.~\eqref{Eq:IsGeneral}, $\boldsymbol{f}_s=\boldsymbol{0}$ because the NM acts as a perfect spin sink for the spin current pumped into the metallic layer.

\begin{figure}[ht] 
\centering 
\includegraphics[scale=1.0]{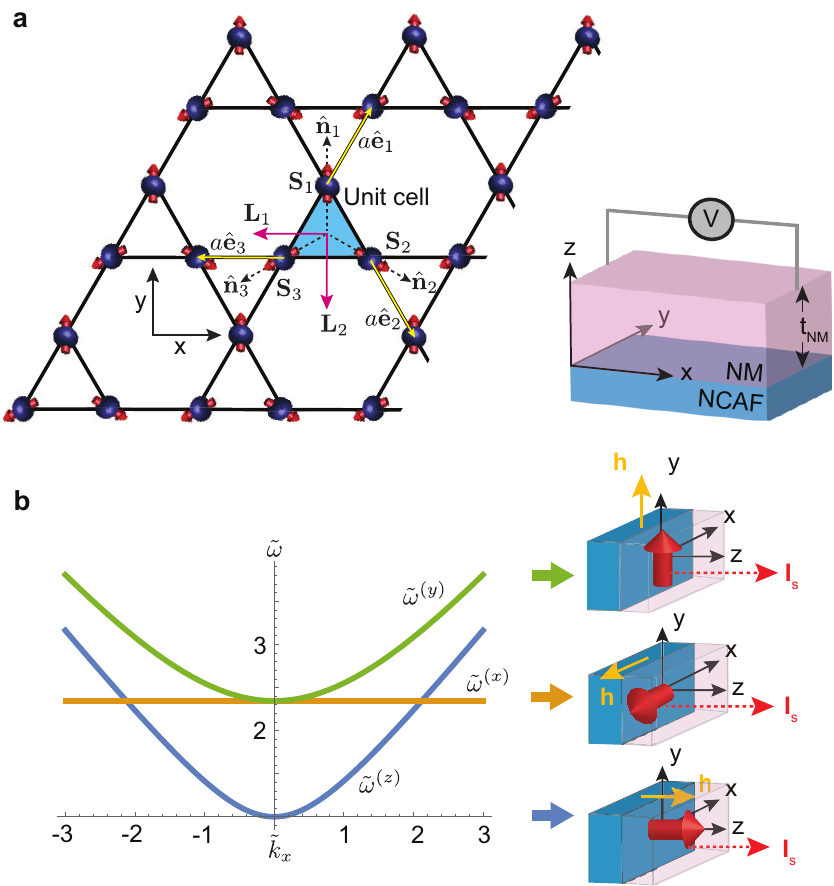}  
\caption{(color online). {\bf a}.~A kagome AF at equilibrium. The spin order is characterized by the staggered fields $\boldsymbol{L}_1= (\boldsymbol{S}_3 - \boldsymbol{S}_2)/\sqrt{3}S$,
$\boldsymbol{L}_2= ( \boldsymbol{S}_2 + \boldsymbol{S}_3 - 2\boldsymbol{S}_1)/3S$.~\cite{Comment1} The thin-film kagome AF is interfaced with a NM along $z$.
{\bf b}.~The spin-wave dispersion relation $\tilde{\omega}^{(i)}= (\tilde{\omega}^{(i)2}_0 + c^{(i)} \tilde{k}_x^2 )^{1/2}$ with $K_z/K = 10$. We have introduced the dimensionless quantities $c^{(x)}=0$, $c^{(y)}=c^{(z)}=1$,
$\tilde{\omega}^{(i)}= \omega^{(i)}/\omega^{(z)}_0$, $\tilde{k}_x= k_x ( a \sqrt{3} JS/\hbar \omega^{(z)}_0)$.
For a kagome AF resonantly driven at frequencies $\tilde{\omega}^{(i)}_0$ by $\boldsymbol{h}$ (yellow arrows), 
the three spin-wave bands pump spin currents with mutually orthogonal spin polarizations (red arrows).   }
\label{Fig2} 
\end{figure} 

\section{Spin pumping in kagome lattices} \label{Sec:Kagome}
We now apply the general theory to a thin-film (monolayer) NCAF with kagome structure~\cite{Comment3} 
interfaced with a NM along $z$ (see Fig.~\ref{Fig2}a). 
Important examples of kagome AFs include Weyl semimetals and iron jarosites~\cite{Kuroda:nm2017,Matan:prl2006}. 
The Hamiltonian of the spin system is~\cite{Ulloa:prb2016,Rodrigues:arxiv21}
\begin{equation}
H = J\sum_{\langle ij \rangle} \vect{S}_i\cdot\vect{S}_j +  \sum_{i}  \left[ K_z \left( \vect{S}_i\cdot\hat{\boldsymbol{e}}^{(z)} \right)^2 - K\left( \vect{S}_i\cdot\hat{\vect{n}}_i \right)^2 \right]. \label{Eq:HamiltonianKagome}
\end{equation}
Here, the first term represents the nearest-neighbor exchange interaction $J>0$, whereas the last term is 
the anisotropy energy with $K_z>0$ and $K>0$. 
$\hat{\boldsymbol{e}}^{(z)}$ is the unit vector along $z$.
Note that the three magnetic sublattices of the kagome AF experience different in-plane easy axes defined by   
$\hat{\vect{n}}_1 = [0,1,0]$, $\hat{\vect{n}}_2 = [\sqrt{3}/2,-1/2,0]$, and $\hat{\vect{n}}_3 = [-\sqrt{3}/2,-1/2,0]$. 
Consequently, the ground state of the kagome AF is given by a $120^{\circ}$ ordering of the sublattice spins such that $\vect{S}_i =  S \hat{\vect{n}}_i$ (or $\vect{S}_i =  - S \hat{\vect{n}}_i$).

Following Ref.~\onlinecite{Dombre:prb1989}, the Lagrange density~\eqref{Eq:Lgeneral} of the above spin Hamiltonian can be microscopically derived (see Sec.~\ref{Sec:Appendix}).
For the constants in Eq.~\eqref{Eq:Lgeneral}, we find that $a_1= 24\hbar S/ \sqrt{3} a^2$ and $a_2= 36 S^2 J / \sqrt{3}a$, where $a$ is the lattice constant.
The exchange energy tensor has the non-vanishing tensor elements $ \Gamma_{xx}^{yy} = \Gamma_{yy}^{xx}= \Gamma_{zz}^{xx}= \Gamma_{zz}^{yy}= 4\sqrt{3} J S^2/a$ and $\Gamma_{xy}^{xy}= \Gamma_{xy}^{yx}= - 4\sqrt{3} J S^2/a$, whereas for the second rank tensors we find $\kappa_{xx}= \kappa_{yy}= K_1$, $\kappa_{zz}= K_2$, $\tilde{\kappa}_{xx}=\tilde{\kappa}_{yy}= a_2$, and $\tilde{\kappa}_{zz}= a_2 + 4\sqrt{3} K_z S^2/a$. 
Here, we have introduced the anisotropy constants $K_1= 8\sqrt{3} (K_z + K) S^2 / a^3$ and $K_2= 16\sqrt{3} K S^2 / a^3$.
Eq.~\eqref{Eq:NonLinSigma} implies that the anisotropic part of  $\tilde{\kappa}_{ij}$ yields a correction on the order of $\sim K_z/J$ to the spin dynamics.
In AFs, the exchange energy is typically much larger than the anisotropy energy.
In what follows, we therefore disregard the anisotropic part and assume $\tilde{\kappa}_{ij}= a_2\delta_{ij}$.
The thermodynamic forces in the spin pumping expression~\eqref{Eq:IsGeneral} are $\boldsymbol{f}_r(t)= -2V\boldsymbol{\mathcal{K}}\cdot\boldsymbol{r}(t)$ and $\boldsymbol{f}_m= -2 a_2 V \boldsymbol{m} (t) + V\boldsymbol{h} (t)$.  Here, $\boldsymbol{\mathcal{K}}$ is a diagonal matrix with $\mathcal{K}_{xx}=\mathcal{K}_{yy}=K_{1}$ and $\mathcal{K}_{zz}=K_{2}$.
To find the time-dependence of $\boldsymbol{f}_r$ and $\boldsymbol{f}_m$, we solve the equations of motion for $\boldsymbol{r}$ and $\boldsymbol{m}$ in the linear response regime.
With $\boldsymbol{f}_s= \boldsymbol{0}$, Eq.~\eqref{Eq:OnsagerMatrix} yields 
$2a_2\boldsymbol{m}= a_1\dot{\boldsymbol{r}} + \boldsymbol{h}$ and $a_1^2\ddot{\boldsymbol{r}} + 4a_2 \boldsymbol{\mathcal{K}}\cdot\boldsymbol{r}= -a_1\dot{\boldsymbol{h}} - 4a_2\tilde{\alpha}\dot{\boldsymbol{r}}$.
Substitution of the ansatz $\boldsymbol{h}(t)=\mathcal{R}e[\boldsymbol{h}_{0}e^{i\omega t}]$ and $\boldsymbol{r}(t)=\mathcal{R}e[\boldsymbol{r}_{0}e^{i\omega t}]$ into the equation for $\boldsymbol{r}$, produces the stationary solutions
\begin{align}
    r_{i}&=-h_{0,i}\Gamma[L_{s}^{(i)}(\omega)\cos{(\omega t)}+L_{a}^{(i)}(\omega)\sin{(\omega t)}] , \\ 
    m_{i}&=\frac{h_{0,i} \mathcal{A}^{(i)} }{2a_2} \bigg[L_{s}^{(i)}(\omega)\sin{(\omega t)}-L_{a}^{(i)}(\omega)\cos{(\omega t)}\bigg] , \label{Eq:mSol}
\end{align}
where $\mathcal{A}^{(i)} = 4\Gamma (\hat{\boldsymbol{e}}^{(i)}\cdot \boldsymbol{\mathcal{K}}\cdot\hat{\boldsymbol{e}}^{(i)}) a_2 / a_1 \omega_0^{(i)} $ ($\hat{\boldsymbol{e}}^{(i)}$ are the three unit vectors along the $x$, $y$, and $z$ axes, respectively), $\Gamma=1/2a_{1}\Delta\omega$ with $\Delta\omega= 2 a_2 \tilde{\alpha}/a_1^2$, and $\omega^{(x)}_0=\omega^{(y)}_0=\sqrt{4K_{1}a_{2}/a_{1}^{2}}$, $\omega^{(z)}_0=\sqrt{4K_{2}a_{2}/a_{1}^{2}}$ are the resonance frequencies for the three spin wave bands of the kagome AF (i.e., the frequencies of the $\boldsymbol{k}=\boldsymbol{0}$ spin waves). 
$L_{s}^{(i)}(\omega)=(\Delta\omega)^{2}/((\delta\omega^{(i)})^{2}+(\Delta\omega)^{2})$ and $L_{a}^{(i)}(\omega)=\delta\omega^{(i)}\Delta\omega/((\delta\omega^{(i)})^{2}+(\Delta\omega)^{2})$ are symmetric and antisymmetric functions of $\delta\omega^{(i)}=\omega-\omega^{(i)}_0$, respectively. 
In arriving at the above expressions, we have expanded around the resonance frequencies and assumed $\Delta\omega/\omega^{(i)}\ll1$. 
Substituting the above stationary solutions into the thermodynamic forces in Eq.~\eqref{Eq:IsGeneral}, we arrive at an expression for the pumped spin current~\cite{Comment2}     
\begin{equation}\label{spin_current_linear}
   I_{s,i}=h_{0,i}[g_1^{(i)} (\omega)\sin{(\omega t)} - g_2^{(i)}(\omega) \cos{(\omega t)} ] . 
\end{equation}
Here, the frequency-dependent functions are 
$g_{1}^{(i)} (\omega) =  V\mathcal{A}^{(i)} [\tau L_{s}^{(i)}(\omega)/a_1-\lambda \omega_0^{(i)}L_{a}^{(i)}(\omega)/2a_2]$ and 
$g_{2}^{(i)} (\omega) = V\mathcal{A}^{(i)} [\lambda \omega_0^{(i)}L_{s}^{(i)}(\omega)/2a_2 + \tau L_{a}^{(i)}(\omega)/a_1]$.
Eq.~\eqref{spin_current_linear} is the second central results of this Letter and provides a theory for the spin pumping of NCAFs with kagome structure. 
Importantly, we notice that the spin current reduces to $I_{s,i}=V\mathcal{A}^{(i)}h_{0,i}[ \tau \sin{(\omega^{(i)}_0 t)}/a_1- \lambda\omega_0^{(i)} \cos{(\omega^{(i)}_0 t)}/2a_2]$ at the resonance frequencies $\omega^{(i)}_0$ where $L_{s}^{(i)}=1$ and $L_{a}^{(i)}=0$. This implies that the reactive (dissipative) STT parameter can be extracted from the in-phase (quadrature) component of $\boldsymbol{I}_s$ with respect to the driving field. 

By applying the driving field along $\hat{\boldsymbol{e}}^{(i)}$, the spin current peaks at the resonance frequency $\omega^{(i)}_0$. The induced current then only contains contributions from one of the three spin wave bands -- the $\boldsymbol{k}=\boldsymbol{0}$ spin wave with a frequency of $\omega^{(i)}_0$.
Interestingly, this makes it very easy to change the polarization direction of the pumped spin current since the different spin wave bands lead to different polarizations (Fig.~\ref{Fig2}b). 
Specifically, a driving field $\boldsymbol{h} = h_0\hat{\boldsymbol{e}}^{(i)} \cos  ( \omega^{(i)}_0 t )$ generates a spin current with a spin polarization along $\hat{\boldsymbol{e}}^{(i)}$.
This is very different from the situation in ferromagnets and collinear AFs, where the entire magnetic state must be rotated for changing the polarization direction of the pumped spin current. 

A common way to detect spin pumping is to interface the magnet with a NM having a large spin Hall angle. 
The spin current injected into the normal NM-layer generates a transverse charge current due to the ISHE, which produces a measurable Hall voltage~\cite{Jiao:prl2013,Saitoh:apl2006,Mosendz:prb2010}. 
We assume the NM-layer to be a heavy metal with strong SOC such that the backflow of spin into the magnetic layer can be neglected. 
Further, we consider the thickness $t_{NM}$ of the NM to be much larger than the spin diffusion length $\lambda_{sd}$ so that the spin current vanishes completely at the outer edge where $z= t_{NM}$. 
An external magnetic field $\boldsymbol{h}=\boldsymbol{h}_{0}\cos{(\omega t)}$ is used to excite the magnet.
A spin current $\boldsymbol{I}_s $ (given by Eq.~\eqref{spin_current_linear}) is then pumped into the normal metal. 
The spin current density $\boldsymbol{j}_{s}$ through the NM layer is found by solving the spin-diffusion equation $\partial_t \boldsymbol{\rho}_s = D\partial_z^2 \boldsymbol{\rho}_s - \boldsymbol{\rho}_s/\tau_{sf}$ with the boundary condition 
$-D\partial_z \boldsymbol{\rho}_s (0,t)= \boldsymbol{I}_s/A$ at the NCAF/NM interface and $\partial_z \boldsymbol{\rho}_s (t_{NM},t)= \boldsymbol{0}$ at the outer edge.
Here, $\boldsymbol{\rho}_s$ is the spin density in the normal metal, $D$ is the electron diffusion constant, $\tau_{sf}$ is the spin-flip time, and $A$ is the cross section area of the NCAF/NM interface.
The spin current density is found from the solution of the spin density via the relationship $\boldsymbol{j}_{s} = -D \partial_z \boldsymbol{\rho}_s$, which yields ~\cite{Mosendz:prb2010}
$\boldsymbol{j}_{s}(z,t)= -(\boldsymbol{I}_s/A)  \sinh{[(z-t_{NM})/\lambda_{sd}]}  /  \sinh{(t_{NM}/\lambda_{sd})}$ (here, $\lambda_{sd}= \sqrt{D\tau_{sf}}$).
The charge current density generated by the ISHE is \cite{Jiao:prl2013,Mosendz:prb2010} $\boldsymbol{j}^{ISHE}_{c}=\gamma_{H}(2e/\hbar) [\hat{\boldsymbol{e}}^{(z)}\times\boldsymbol{j}_{s}]$,
where $\gamma_{H}$ is the spin Hall angle. The system constitutes an open circuit. Thus, the deflected charges accumulate at the interfaces, which induce an electric field that exactly 
cancel $\boldsymbol{j}^{ISHE}_{c}$. Integrating the net current density over the metallic layer, one finds the electric field~\cite{Mosendz:prb2010},  
\begin{equation}
    \boldsymbol{E}_{ISHE}=-\frac{2e\gamma_{H}\lambda_{sd} }{\hbar A\sigma_{NM}t_{NM}}\tanh{\bigg(\frac{t_{NM}}{2\lambda_{sd}}\bigg)}[\hat{\boldsymbol{e}}^{(z)}\times \boldsymbol{I}_{s} ],
\end{equation}
where $\sigma_{NM}$ is the conductance of the NM-layer. We have here disregarded any electric currents in the kagome AF layer because it is a thin film. 
$\boldsymbol{E}_{ISHE}$ is proportional to the spin current. Thus, the reactive (dissipative) STT can be determined from the in-phase (quadrature) component of the electric signal with respect to $\boldsymbol{h}(t)$. 

\section{Conclusion}\label{Sec:Conclusions}
In conclusion, we have derived a general theory for spin pumping in NCAFs and applied the formalism to NCAFs with kagome lattice. Our findings reveal that spin pumping represents a powerful mechanism for exploring both the reactive and dissipative STTs of NCAFs. We show that the reactive (dissipative) part of the STT is proportional to the in-phase (quadrature) component of the pumped spin current at resonance. Additionally, we find that the three spin-wave bands of the kagome AF lead to currents with spin polarizations along the $x$, $y$, and $z$ axis, respectively. This makes it possible to orient the spin current along any axis by exciting different spin-wave modes. Thus, our work demonstrates that the spin pumping of NCAFs is richer and more complex than in ferromagnets and collinear AFs, and opens the door for tuning the spin current's orientation through the frequency of $\boldsymbol{h}$.

\section{Acknowledgements}
This work received funding from the Research Council of Norway via the Young Research Talents Grant No. 286889 "Antiferromagnetic Spinmechatronics".

\appendix
\section{Effective action of kagome AFs}\label{Sec:Appendix}
The kagome AF is modeled by the spin Hamiltonian~\eqref{Eq:HamiltonianKagome}, where the three magnetic sublattices are connected by the vectors $\hat{\boldsymbol{e}}_{1}=[1/2,\sqrt{3}/2,0]$, $\hat{\boldsymbol{e}}_{2}=[1/2,-\sqrt{3}/2,0]$ and $\hat{\boldsymbol{e}}_{3}=[-1,0,0]$ (see Fig.~\ref{Fig2}a).
The order parameter of the spin system is a rotation matrix $\boldsymbol{R}\in SO(3)$, which defines the local orientation of the reference frame spanned by the mutually orthogonal staggered fields.~\cite{Comment1} 
Additionally, we introduce a vector field $\boldsymbol{L}$ representing a tilting of the spins $\boldsymbol{S}_{1}$, $\boldsymbol{S}_{2}$, and $\boldsymbol{S}_{3}$. $\boldsymbol{R}$ and $\boldsymbol{L}$ both have three degrees of freedom each. Thus, $\boldsymbol{R}$ and $\boldsymbol{L}$ can parametrize all possible configurations of the three sublattice spins in a unit cell. The spin on the $i$th sublattice can therefore be expressed as
\begin{equation}
    \boldsymbol{S}_{i}=\frac{S\boldsymbol{R}(\hat{\boldsymbol{n}}_{i}+a\boldsymbol{L})}{\sqrt{1+2a\hat{\boldsymbol{n}}_{i}\cdot\boldsymbol{L}+a^2\boldsymbol{L}^2}}. \label{Eq:Sgeneral}
\end{equation}
Here, $a$ is the lattice constant and the denominator is introduced to ensure that the spin vectors are correctly normalized. 
We consider an AF with large exchange energy, which implies that $|a\boldsymbol{L}|\ll1$.

The action governing the spin dynamics is given by
\begin{equation}\label{General_action}
    \mathcal{S}=\sum_{i}\int dt\mathcal{L}_{i},
\end{equation}
where $\mathcal{L}_{i}=\mathcal{T}_{i}-\mathcal{U}_{i}-\mathcal{U}_{s,i}$ is the Lagrangian density of the spin at lattice site $i$. The first term in $\mathcal{L}_{i}$ describes the kinetic energy and is given by $\mathcal{T}_{i}=\hbar\boldsymbol{A}(\boldsymbol{S}_{i})\cdot \dot{\boldsymbol{S}}_{i}$, where $\boldsymbol{A}$ is a vector potential satisfying $\nabla\times\boldsymbol{A}(\boldsymbol{S}_{i})=\boldsymbol{S}_{i}/S$. $\mathcal{U}_{i}=H_{i}-g\boldsymbol{B}\cdot\boldsymbol{S}_{i}$ represents the interaction energy, where we have included the coupling to an external magnetic field $\boldsymbol{B}$. $H_i$ is the contribution of lattice site $i$ to the Hamiltonian~\eqref{Eq:HamiltonianKagome}. The coupling to the spin accumulation is $\mathcal{U}_{s,i}=\hbar\lambda_{r}\boldsymbol{f}_{s}\cdot\boldsymbol{S}_{i}$, where $\lambda_{r}$ parametrizes the reactive STT.~\cite{Maekawa:book}  Because we consider a thin-film NCAF, $\boldsymbol{f}_s$ is constant and determined by the spin accumulation at the NM/NCAF interface. 
Here, and below in the case of the dissipative STT, we disregard effects of the SOC that breaks the spin rotational symmetry of the STT.

In the following, we will derive an effective action describing the low-frequency, long-wavelength dynamics of the kagome AF. To this end, we follow Ref.~\onlinecite{Dombre:prb1989} and expand the action to second order in
the external force fields $\{\boldsymbol{B},\boldsymbol{f}_{s}\}$ and out-of-equilibrium quantities
(i.e., $\partial_{\mu}R$ ($\mu=\{t,x,y,z\}$) and $a\boldsymbol{L}$). 
Consequently, we only require terms up to first order in $a\boldsymbol{L}$ in the spin vector~\eqref{Eq:Sgeneral} and use the approximation: 
\begin{equation}\label{S_approx}
    \boldsymbol{S}_{i}\approx S\boldsymbol{R}(\hat{\boldsymbol{n}}_{i}+\boldsymbol{\Delta}_{i}), \; \; \; \; \; \; \; \; \;(\boldsymbol{\Delta}_{i}\cdot\hat{\boldsymbol{n}}_{i}=0) .
\end{equation}
Here, $\boldsymbol{\Delta}_{i}=a(\boldsymbol{L}-(\hat{\boldsymbol{n}}_{i}\cdot\boldsymbol{L})\hat{\boldsymbol{n}}_{i})$. The total spin polarization of a unit cell is then $\boldsymbol{S}_{tot}=\sum_{k=1}^{3}\boldsymbol{S}_{k}=3aS\boldsymbol{R}(\boldsymbol{T}\boldsymbol{L})$, where we have introduced the operator $T_{\alpha\beta}=\delta_{\alpha\beta} - (1/3)\sum_{k=1}^{3} n_{k,\alpha}n_{k,\beta}$. In our case, the operator is diagonal with elements $2T_{xx}=2T_{yy}=T_{zz}=1$.

To formulate a continuum model of the action in Eq.~\eqref{General_action}, it is convenient to first consider the energy contribution from one unit cell (by grouping the three spins in each unit cell) and then sum over all unit cells. We follow this approach for all the calculations below. 

First, we consider the kinetic energy term. The kinetic energy of one unit cell is given by $\mathcal{T}^{(\Delta)}=\sum_{k}\hbar A_{\alpha}[\boldsymbol{S}_k] \dot{{S}}_{k,\alpha}$, where $k\in \{1,2,3\}$ labels the three spins in the unit cell and Einstein's summation convention is implied for the $\alpha$-index. Using Eq.~\eqref{S_approx} and expanding the vector potential $\boldsymbol{A}(\boldsymbol{S}_{k})$ to first order in the out-of-equilibrium quantities, one finds
\begin{eqnarray}\label{Eq:Texp}
\sum_{k}\hbar A_{\alpha}[\boldsymbol{S}_k]  \dot{{S}}_{k,\alpha} &\approx& \sum_{k}\hbar S[A_{\alpha}(\boldsymbol{R}\hat{\boldsymbol{n}}_k)\cdot (\dot{\boldsymbol{R}}\hat{\boldsymbol{n}}_k)_{\alpha}+\nonumber \\
& & a\epsilon_{\alpha\beta\gamma}L_{\alpha}n_{k,\beta}(\boldsymbol{R}^{T} \dot{\boldsymbol{R}}\hat{\boldsymbol{n}}_k)_{\gamma}] .
\end{eqnarray}
Here, we have (in the last term) utilized the relationship between the vector potential and spin vector, and the property $\epsilon_{\alpha\beta\gamma}\boldsymbol{R}_{\alpha\alpha^{'}}\boldsymbol{R}_{\beta\beta^{'}}\boldsymbol{R}_{\gamma\gamma^{'}}=\epsilon_{\alpha^{'}\beta^{'}\gamma^{'}}$ of the rotation matrix. The first term in Eq.~\eqref{Eq:Texp} is a topological term~\cite{Dombre:prb1989} that does not affect the dynamics. Therefore, we disregard this term in what follows. Because the rotation matrix is orthogonal (i.e., $\boldsymbol{R}^{T}\boldsymbol{R}=\boldsymbol{I}$), the quantity $\boldsymbol{R}^{T}\dot{\boldsymbol{R}}$ is antisymmetric. It can therefore be written as $(\boldsymbol{R}^{T}\dot{\boldsymbol{R}})_{ij}=-\epsilon_{ij\alpha}\mathcal{V}_{\alpha}$, where $\mathcal{V}_{x}$, $\mathcal{V}_{y}$, and $\mathcal{V}_{z}$ parameterize the three independent matrix elements.       
Using this expression and summing over the three spins, the kinetic energy of one unit cell becomes $\mathcal{T}^{(\Delta)}=3a\hbar S(\boldsymbol{T}\boldsymbol{L})\cdot\boldsymbol{\mathcal{V}}$.  

Second, we consider interaction energy $\mathcal{U}$. We start with the Heisenberg exchange term $H=J\sum_{\langle ij\rangle}\boldsymbol{S}_{i}\cdot\boldsymbol{S}_{j}$, where the energy contribution from one unit cell is 
\begin{eqnarray}\label{exchange}
    H^{(\Delta)}_{ex} &=& J[\boldsymbol{S}^{l}_{1}\cdot(\boldsymbol{S}^{l+\hat{e}_{1}}_{3}+\boldsymbol{S}^{l-\hat{e}_{1}}_{3})+\boldsymbol{S}^{l}_{2}\cdot(\boldsymbol{S}^{l+\hat{e}_{2}}_{1}+ \nonumber \\ 
  & &  \boldsymbol{S}^{l-\hat{e}_{2}}_{1})+\boldsymbol{S}^{l}_{3}\cdot(\boldsymbol{S}^{l+\hat{e}_{3}}_{2}+\boldsymbol{S}^{l-\hat{e}_{3}}_{2})].
\end{eqnarray}
Here, $l$ denotes the position of the spin in the unit cell, whereas $l\pm \hat{e}_{1}$ is the neighboring lattice site connected to $l$ via the lattice vector $\pm a\hat{\boldsymbol{e}}_i$ (see Fig.~\ref{Fig2}a). Substituting the gradient expansion $\boldsymbol{S}^{l\pm\hat{e}_{i}}_{j}\approx\boldsymbol{S}^{l}_{j}\pm a(\hat{\boldsymbol{e}}_{i}\cdot\boldsymbol{\nabla})\boldsymbol{S}_{j}^{l}+\frac{a^{2}}{2}(\hat{\boldsymbol{e}}_{i}\cdot\boldsymbol{\nabla})^{2}\boldsymbol{S}_{j}^{l}$ along with the expression \eqref{S_approx} into Eq.~\eqref{exchange}, yield to second order in $a\boldsymbol{L}$ and the spatial gradients of $\boldsymbol{R}$ the following exchange energy of a unit cell
\begin{equation}
    H^{(\Delta)}_{ex}=9a^{2}S^{2}J(\boldsymbol{T}\boldsymbol{L})^{2}+ V_c\Lambda_{ij}^{\alpha\beta}[\partial_{\alpha}\boldsymbol{R}^{T}\partial_{\beta}\boldsymbol{R}]_{ij}.
\end{equation}
The tensor in the last term is defined as $\Lambda_{ij}^{\alpha\beta}= -(4S^{2}J/a\sqrt{3})[n_{1,i}n_{3,j}e_{1,\alpha}e_{1,\beta}+n_{2,i}n_{1,j}e_{2,\alpha}e_{2,\beta}+n_{3,i}n_{2,j}e_{3,\alpha}e_{3,\beta}]$,
where $V_c=a^3\sqrt{3}/4$. 

The in-plane anisotropy energy per unit cell is given by $H^{(\Delta)}_{in}=-\sum_{k}K(\boldsymbol{S}_{k}\cdot\hat{\boldsymbol{n}}_{k})^{2}$. Using Eq. \eqref{S_approx}, we find to second order in the out-of-equilibrium quantities 
\begin{equation}
    H^{(\Delta)}_{in}=-KS^{2}\sum_{\gamma}(n_{\gamma, i}n_{\gamma, j}n_{\gamma, k}n_{\gamma, l})(R_{ij}R_{kl}) . 
\end{equation}
Here, we have made the substitution $k\rightarrow\gamma$ for the summation index to get agreement with the indices used in Eq.~\eqref{Eq:Ugeneral}. 
Similarly, we find for the out-of-plane anisotropy energy $H^{(\Delta)}_{out}=\sum_{k}K_{z}(\hat{\boldsymbol{z}}\cdot\boldsymbol{S}_{k})^2$ the expression
\begin{equation}
    H^{(\Delta)}_{out}=K_{z}S^{2}\sum_{\gamma}n_{\gamma, i}n_{\gamma, j}R_{zi}R_{zj}+3a^{2}S^{2}K_{z}(\hat{\boldsymbol{z}}\cdot \boldsymbol{T}\boldsymbol{L})^{2}.
\end{equation}
The potential energy associated with the coupling of the spins to an external magnetic field is given by the Zeeman energy $H^{(\Delta)}_{B}=-\sum_{k}g\boldsymbol{B}\cdot\boldsymbol{S}_{k}$. 
Considering the field to be spatially uniform, $H^{(\Delta)}_{B}$ to second order in the external force field and out-of-equilibrium quantities is
\begin{equation}
    H^{(\Delta)}_{B}=-3gaS\boldsymbol{B}\cdot(\boldsymbol{T}\boldsymbol{L}),
\end{equation}
where we have used that $\sum_{k}\boldsymbol{S}_{k}=3aS\boldsymbol{R}(\boldsymbol{T}\boldsymbol{L})$. Thus, the total interaction energy of a unit cell becomes 
$\mathcal{U}^{(\Delta)}= H^{(\Delta)}_{ex} + H^{(\Delta)}_{in} + H^{(\Delta)}_{out} + H^{(\Delta)}_{B}$.

Lastly, we consider the interaction energy $\boldsymbol{\mathcal{U}}_{s}^{(\Delta)}=\sum_{k}\hbar\lambda_{r}\boldsymbol{f}_{s}\cdot\boldsymbol{S}_{k}$ (per unit cell) produced by a spin accumulation.
We disregard the spatial variations in the spin accumulation. To second order in $\boldsymbol{f}_{s}$ and the out-of-equilibrium quantities, one then finds  
\begin{equation}
    \mathcal{U}_{s}^{(\Delta)}= 3\hbar aS\lambda_{r}\boldsymbol{f}_{s}\cdot(\boldsymbol{T}\boldsymbol{L}). 
\end{equation}

Combining the interaction terms and summing over all unit cell, the action functional becomes $\mathcal{S}=\int dt\sum_{\Delta}\mathcal{L}^{(\Delta)}$. Here,
$\mathcal{L}^{(\Delta)}= \mathcal{T}^{(\Delta)} - \mathcal{U}^{(\Delta)} - \mathcal{U}_{s}^{(\Delta)}$ is the Lagrangian density of a unit cell.
In order to obtain the continuous action, we take the continuum limit $\sum_{\Delta}\rightarrow\int \frac{dxdydz}{V_{c}}$. Since we consider a monolayer with a thickness of $L_z\sim a$, the constant $V_c = aa_c$ where $a_{c}=a^{2}\sqrt{3}/4$ denotes the area of the unit cell of the 2D kagome lattice. Thus, we obtain the action $\mathcal{S}=\int dVdt\mathcal{L}=\int dVdt(\mathcal{T}-\mathcal{U}-\mathcal{U}_{s})$, where the kinetic, interaction, and STT energies are given by 
\begin{align}
    \mathcal{T}&=\frac{a_{1}}{2}\boldsymbol{\mathcal{V}}\cdot\boldsymbol{m},\label{Kinetic}\\
    \mathcal{U}&=\Lambda^{\alpha\beta}_{ij}[\partial_{\alpha}\boldsymbol{R}^{T}\partial_{\beta}\boldsymbol{R}]_{ij}+\nu^{kl}_{ij}R_{ij}R_{kl}+\Tilde{\kappa}_{ij}m_{i}m_{j}-\boldsymbol{h}\cdot\boldsymbol{m},\label{interaction}\\
    \mathcal{U}_{s}&=\lambda\boldsymbol{m}\cdot\boldsymbol{f}_{s}.\label{spin_coupling}
\end{align}
The kinetic term is parametrized by the constant $a_{1}=24\hbar S/a^2\sqrt{3}$, and we have introduced the vector field $\boldsymbol{m}=\boldsymbol{T}\boldsymbol{L}$. Furthermore, we have defined the tensors $\nu_{ij}^{kl}=(4S^2/a^{3}\sqrt{3})\sum_{\gamma=1}^{3}(K_{z}n_{\gamma, j}n_{\gamma, l}\delta_{z,i}\delta_{z,k}-Kn_{\gamma, i}n_{\gamma, j}n_{\gamma, k}n_{\gamma, l})$ and  $\Tilde{\kappa}_{ij}=a_{2}\delta_{ij}+\eta_{ij}$, with $a_{2}=36S^{2}J/a\sqrt{3}$ and $\eta_{ij}= (12K_{z}S^{2}/a\sqrt{3})\delta_{zi}\delta_{zj}$. The vector $\boldsymbol{h}$  is related to the external field $\boldsymbol{B}$ by $\boldsymbol{h}=12 g S\boldsymbol{B}/a^2\sqrt{3}$, and the STT coupling parameter is $\lambda=12\hbar S\lambda_{r}/a^2\sqrt{3}$.

\section{Dissipation functional}\label{Sec:Dissipation}
The dissipative processes of the spin system is captured by the Rayleigh dissipation functional~\cite{Maekawa:book}
\begin{equation}\label{General_diss}
    \mathcal{G}=\sum_{i}\int dt\bigg(\frac{\hbar\alpha_{G}}{2}(\dot{\boldsymbol{S}}_{i})^2+\hbar\lambda_{d}\dot{\boldsymbol{S}}_{i}\cdot(\boldsymbol{f}_{s}\times\boldsymbol{S}_{i})\bigg) ,
\end{equation}
where $\alpha_{G}$ is the Gilbert damping and $\lambda_{d}$ determines the dissipative STT. 
Note that works on collinear AFs have shown that additional cross-sublattice dissipative processes could play a role in cases where the AF/NM interface breaks the sublattice symmetry.~\cite{Kamra:prl2017,Troncoso:prb2021,Kamra:prb2018} 

We first consider the damping term in the dissipation functional. Substituting Eq. \eqref{S_approx} for the spin, we find to second order in the out-of-equilibrium quantities $(\dot{\boldsymbol{S}}_{i})^{2}=\dot{R}_{\alpha\alpha^{'}}\dot{R}_{\alpha\beta^{'}}n_{i\alpha^{'}}n_{i\beta^{'}}$. Summing over the three spins of the unit cell, the energy dissipation associated with the Gilbert damping becomes
\begin{equation}
    \mathcal{G}^{\Delta}_{damp}= \frac{3\hbar\alpha_{G}S^2}{4}\text{Tr}[\dot{\boldsymbol{R}}^{T}\dot{\boldsymbol{R}}]. 
\end{equation}
In arriving at this expression, we have used that $\sum_{k}n_{k\alpha^{'}}n_{k\beta^{'}}=(3/2)\delta_{\alpha^{'}\beta^{'}}$ when the 120 degree ordering is not restricted to lie in the $xy$-plane (which is valid in the limit of vanishing intrinsic SOC). 

The dissipative STT yields to second order in $\boldsymbol{f}_{s}$ and $\dot{\boldsymbol{R}}$ the following contribution to the dissipation functional
\begin{equation}
    \mathcal{G}^{\Delta}_{STT}\approx 3\hbar S^2\lambda_{d} \boldsymbol{\mathcal{V}}\cdot\boldsymbol{f}_{s}. 
\end{equation}
Here, we have used that $\boldsymbol{S}_{k}\times\dot{\boldsymbol{S}}_{k}\approx S^2 (\boldsymbol{R}\boldsymbol{n}_{k})\times(\dot{\boldsymbol{R}}\boldsymbol{n}_{k})$ and summed 
over the three sublattice spins.
Adding the contributions from the two dissipative processes and taking the continuum limit lead to 
\begin{equation}\label{Continuum_diss}
   \mathcal{G}=\int dVdt\bigg(\frac{\Tilde{\alpha}}{8}\text{Tr}[\dot{\boldsymbol{R}}^{T}\dot{\boldsymbol{R}}]+\frac{\Tilde{\lambda}}{2}\boldsymbol{\mathcal{V}}\cdot\boldsymbol{f}_{s}\bigg),
\end{equation}
where $\Tilde{\alpha}=24\hbar\alpha_{G}S^{2}/a^3\sqrt{3}$ and $\Tilde{\lambda}=24\hbar\lambda_{d}S^{2}/a^{3}\sqrt{3}$. 
The action~\eqref{Kinetic}-\eqref{spin_coupling} and dissipation functional~\eqref{Continuum_diss} provide an effective description of the kagome AF. 
Expressing the rotation matrix in terms of the Gibbs vector using Eqs.~\eqref{Eq:GibbsRel1}-\eqref{Eq:GibbsRel2}, one arrives at Eqs.~\eqref{Eq:Lgeneral}-\eqref{Eq:Dissipation2}.



\begin{thebibliography}{99}
\bibitem{Comment3}We assume the kagome AF to be a monolayer with a thicknesses of $L_z\sim a$.
\bibitem{Comment2} Note that Eqs.~\eqref{Eq:mSol} and \eqref{spin_current_linear} also contain a term $\sim \boldsymbol{h}_0 \cos (\omega t)$, which does not depend on the functions $L_s^{(i)}$ and $L_a^{(i)}$.
However, close to the resonance frequencies this term is negligible.
\end{thebibliography}
\end{document}